\documentstyle[psfig]{mn}

\newif\ifAMStwofonts

\ifoldfss
  \ifCUPmtlplainloaded \else
    \NewTextAlphabet{textbfit} {cmbxti10} {}
    \NewTextAlphabet{textbfss} {cmssbx10} {}
    \NewMathAlphabet{mathbfit} {cmbxti10} {} 
    \NewMathAlphabet{mathbfss} {cmssbx10} {} 
  \fi
  \ifAMStwofonts
    \ifCUPmtlplainloaded \else
      \NewSymbolFont{upmath} {eurm10}
      \NewSymbolFont{AMSa} {msam10}
      \NewMathSymbol{\upi}     {0}{upmath}{19}
      \NewMathSymbol{\umu}     {0}{upmath}{16}
      \NewMathSymbol{\upartial}{0}{upmath}{40}
      \NewMathSymbol{\leqslant}{3}{AMSa}{36}
      \NewMathSymbol{\geqslant}{3}{AMSa}{3E}

      \let\leq=\leqslant \let\le=\leqslant
       \let\ge=\geqslant
    \fi
  \fi
\fi 

\ifnfssone
  \newmathalphabet{\mathit}
  \addtoversion{normal}{\mathit}{cmr}{m}{it}
  \addtoversion{bold}{\mathit}{cmr}{bx}{it}
  \newmathalphabet{\mathbfit} 
  \addtoversion{normal}{\mathbfit}{cmr}{bx}{it}
  \addtoversion{bold}{\mathbfit}{cmr}{bx}{it}
  \newmathalphabet{\mathbfss} 
  \addtoversion{normal}{\mathbfss}{cmss}{bx}{n}
  \addtoversion{bold}{\mathbfss}{cmss}{bx}{n}
  \ifAMStwofonts
    \ifCUPmtlplainloaded \else
      \UseAMStwoboldmath
      \makeatletter
      \new@mathgroup\upmath@group
      \define@mathgroup\mv@normal\upmath@group{eur}{m}{n}
      \define@mathgroup\mv@bold\upmath@group{eur}{b}{n}
      \edef\UPM{\hexnumber\upmath@group}
      \new@mathgroup\amsa@group
      \define@mathgroup\mv@normal\amsa@group{msa}{m}{n}
      \define@mathgroup\mv@bold\amsa@group{msa}{m}{n}
      \edef\AMSa{\hexnumber\amsa@group}
      \makeatother
      \mathchardef\upi="0\UPM19
      \mathchardef\umu="0\UPM16
      \mathchardef\upartial="0\UPM40
      \mathchardef\leqslant="3\AMSa36
      \mathchardef\geqslant="3\AMSa3E

      \let\leq=\leqslant \let\le=\leqslant
       \let\ge=\geqslant
    \fi
  \fi
\fi 

\ifnfsstwo
  \DeclareMathAlphabet{\mathbfit}{OT1}{cmr}{bx}{it}
  \SetMathAlphabet\mathbfit{bold}{OT1}{cmr}{bx}{it}
  \DeclareMathAlphabet{\mathbfss}{OT1}{cmss}{bx}{n}
  \SetMathAlphabet\mathbfss{bold}{OT1}{cmss}{bx}{n}
  \ifAMStwofonts
    \ifCUPmtlplainloaded \else
      \DeclareSymbolFont{UPM}{U}{eur}{m}{n}
      \SetSymbolFont{UPM}{bold}{U}{eur}{b}{n}
      \DeclareSymbolFont{AMSa}{U}{msa}{m}{n}
      \DeclareMathSymbol{\upi}{0}{UPM}{"19}
      \DeclareMathSymbol{\umu}{0}{UPM}{"16}
      \DeclareMathSymbol{\upartial}{0}{UPM}{"40}
      \DeclareMathSymbol{\leqslant}{3}{AMSa}{"36}
      \DeclareMathSymbol{\geqslant}{3}{AMSa}{"3E}

      \let\leq=\leqslant \let\le=\leqslant
       \let\ge=\geqslant
    \fi
  \fi
\fi 

\ifCUPmtlplainloaded \else
  \ifAMStwofonts \else 
    \def\upi{\pi}
    \def\umu{\mu}
    \def\upartial{\partial}
  \fi
\fi

\title{Non-uniform extinction in young open star clusters}
\author[R. K. S. Yadav and Ram Sagar]
       {R. K. S. Yadav$^{1}$\thanks{E-mail: rkant@upso.ernet.in} and Ram Sagar$^{1,2}$\thanks{E-mail: sagar@upso.ernet.in}\\
        $^{1}$State Observatory, Manora Peak Nainital 263129, India\\
        $^{2}$Indian Institute of Astrophysics, Bangalore 560034, India}
\date{Accepted ---------.
      Received ---------;
      }

\pagerange{\pageref{firstpage}--\pageref{lastpage}}
\pubyear{2001}

\begin{document}

\maketitle

\label{firstpage}

\begin{abstract}
The extinction law and the variation of colour excess with position, luminosity as
well as spectral class in young open star clusters NGC 663, NGC 869, NGC 884, NGC 1502, NGC 1893, NGC 2244, NGC 2264, NGC 6611, Tr 14, Tr 15, Tr 16, Coll 228, Tr 37 and Be 86 have been studied. The difference in the minimum and maximum values of $E(B-V)$ of cluster members has been considered as a measure of the presence of non-uniform gas and dust inside the clusters. Its value ranges from 0.22 to 1.03 mag in clusters under study, which indicates that non-uniform extinction is present in all the clusters. It has been noticed for the first time in NGC 1502 and Tr 37. It is also found that the differential colour excess in open clusters, which may be due to the presence of gas and dust, decreases systematically with the age of clusters indicating that matter is used either in star formation or blown away by hot stars or both. There is no uniformity in the variation of $E(B-V)$ with either position or spectral class or luminosity. Except in Tr 14, all clusters show a random spatial distribution of $E(B-V)$ indicating a random distribution of gas and dust inside the clusters. The $E(B-V)$ value correlates with both luminosity and spectral class only in the case of Coll 228, Tr 16 and Be 86. The members of these clusters at $\lambda$ $\ge$ $\lambda_{R}$ show larger values of colour excess ratios than the normal ones. The value of $E(U-V)$/$E(B-V)$ for most of the cluster members is close to the normal interstellar value of 1.73. However, the colour excess ratios with $E(B-V)$ at $\lambda$$\ge$$\lambda_{J}$ are smaller than the normal value for NGC 663, NGC 869, NGC 884 and NGC 1502 while they are larger for NGC 6611, Coll 228, Tr 16 and Tr 14. Thus there is no uniformity in the relationship of extinction properties amongst the clusters under study.

\end{abstract}


\section{Introduction}
The investigation of interstellar matter in star-forming environments
is a front line research area. One of
the most obvious properties of interstellar matter that can be determined observationally is the extinction law, $i.e.$ the dependence of the extinction properties of the matter on wavelength. This has been particularly fruitful in recent years with the extension of observing capabilities beyond the optical, into
the ultra-violet (UV) and infrared.

In the pre-main sequence evolutionary scenario suggested by Larson (1973) and Hayashi (1970) a
star forms in a cloud and contracts gravitationally until a sufficiently high central
temperature, density and pressure are reached for nuclear reactions to begin. As this stellar
core evolves to main sequence (MS) stability, it continues to accrete the surrounding circumstellar
envelope of gas and dust. Obviously,
not all the circumstellar material becomes part of the star; some fraction is left as a remnant envelope of gas and dust. This view is supported  by the infrared
observations of pre-MS objects in NGC 2264 by Warner, Strom $\&$ Strom (1979) and in the Orion nebula cluster by McNamara (1976).
 Also, not all matter of the molecular cloud is used
in forming stars. The young (age $\leq10^{7}$ years) star clusters thus provide
an ideal opportunity for studying the properties of matter
in different star forming regions containing hot O and B stars.

     The non-uniform extinction in open star clusters of intermediate age ($\leq10^{8}$ yrs)
was studied by Burki (1975) using photoelectric UBV photometric values of early type cluster members.
He found that differential extinction is important not only in the case of
very young ($\sim10^{6}$ yrs) clusters, but also in some clusters older than $5\times10^{7}$ years.
Wallenquist (1975) studied non-uniform extinction in open
clusters using star counts from the Palomar Sky Survey.
 The underlying assumption in this type of investigation is that the observed deficiency of stars is
mainly due to the presence of absorbing matter  between
the observer and the cluster. Sagar (1987) studied interstellar
extinction in 15 open clusters using UBV photoelectric photometric observations
of proper motion cluster members, and found that ten of them show non-uniform
 extinction across the cluster region. Reddish (1967) has
shown that reddening increases with stellar
luminosity in all clusters and associations with ages $\leq10^{5}$ years,
 but only in some objects with ages between $10^{5}$ and $2\times10{^6}$ years, and such relations are not observed in the clusters older
than $2\times10{^6}$ years.
 Bohannan (1975) re-examined the
young clusters data of Reddish (1967) and after identifying foreground stars and applying correct intrinsic colour indices for bright supergiants found no correlation of reddening
with luminosity. Sagar (1987) observed that in some young clusters the variation of colour excess $E(B-V)$ correlated with luminosity, in the sense that
brighter cluster members were more highly reddened.
\begin{table*}
 \centering
 \begin{minipage}{140mm}
 \caption{General information taken from Mermilliod (1995) about the
clusters under study. The Galacto-centric distance, no. of stars used in this
analysis, statistically expected number of field stars in the sample and median
proper motion membership probability of the sample are denoted by $R_{G}$,
$N_{S}$, $N_{f}$ and $P_{M}$ respectively. $R_{G}$ is calculated assuming
galacto-centric distance of the sun as 8.5 kpc. In the source column 1, 2, 3, 4
and 5 denote references Muminov (1983), Cudworth et al. (1993), Marschall, van
Altena $\&$ Chiu (1982), Zhao et al. (1985) and Marschall $\&$ van Altena (1987
) respectively.}
  \begin{tabular}{lccrcccccccl}
\hline
IAU number&Sequence no.&l &b~~~&Distance&log (age)&Radius&$R_{G}$&~~~$N_{S}$&$N_{f}$&$P_{M}$&Source \\
&&(deg)&(deg)&(kpc)&(yrs)&(pc)&(kpc)&&&($\%$)& \\
\hline
C0142 $+$ 610&NGC 663&129.46&  $-0.94$&2.3&7.1&~~5.2&10.1&~166&&& \\
C0215 $+$ 569&NGC 869&134.63&  $-3.72$&2.1&7.1&~~9.6&10.2&~~38&12&68&~~1 \\
C0218 $+$ 568&NGC 884&135.08&  $-3.60$&2.5&7.1&10.0&10.2&~~46&11&78&~~1 \\
C0403 $+$ 622&NGC 1502&143.65& $ 7.62$&0.9&7.0&~~1.0&~~9.2&~~24&&& \\
C0519 $+$ 333&NGC 1893&173.59& $-1.70$&3.9&7.0&~~6.4&12.4&~101&&& \\
C0629 $+$ 049&NGC 2244&206.40& $-2.02$&1.6&6.8&~~6.0&~~9.9&~~81&5&94&~~3 \\
C0638 $+$ 099&NGC 2264&202.94& $ 2.20$&0.8&6.9&~~2.2&~~9.2&~~26&4&86&~~4 \\
C1041 $-$ 593&Tr 14&287.42&    $-0.58$&3.0&6.7&~~2.0&~~8.1&~144&8&91&~~2 \\
C1042 $-$ 591&Tr 15&287.40&    $-0.36$&1.5&6.7&~~1.0&~~8.1&~~20&&& \\
C1043 $-$ 594&Tr 16&287.61&    $-0.65$&3.0&6.4&~~4.2&12.4&~184&5&95&~~2 \\
C1041 $-$ 597&Coll 228&287.52& $-1.03$&2.3&6.7&~~6.0&~~8.1&~118&&& \\
C1816 $-$ 138&NGC 6611&~16.99& $ 0.79$&2.1&6.1&~~2.5&~~5.9&~127&8&88&~~4 \\
C2018 $+$ 663&Be 86&~76.66&    $ 1.26$&1.1&7.1&~~2.0&~~8.3&~~64&&& \\
C2137 $+$ 572&Tr 37&~99.29&    $ 3.73$&0.8&7.2&~~5.5&~~8.7&~~46&7&93&~~5 \\ \hline
\end{tabular}
\end{minipage}
\end{table*}
Most of the above analyses were based on UBV data taken primarily with either photographic plates or single-channel photometers. Current optical and infrared imaging technology allows us to derive more accurate photometry, as the effects of nebular surface brightness which are generally present in young clusters can be properly removed. Recently, such observations have become available
in the literature. We have therefore used them to study the non-uniform extinction as well as its nature in 14 young open clusters. Section 2 describes the selection of the sample. In Section 3 the details of the reddening determinations are given, while the results derived from the present analysis and their discussions are given in the remaining part of the paper.
\section[]{Selection of clusters}
This section provides information about the criteria
adopted for selection of clusters and their
members along with information about observational data.
\subsection{Star clusters and observational data}
The observations used in this study are taken from the compilation database on star clusters by Mermilliod (1995) at web site {\bf http://obswww.unige.ch/webda/}. We have selected those 14 young (age $<$ 20 Myr) open clusters which have JHK data for at least 10 cluster members. The general information about them are listed in Table 1. The distances of the sample clusters range from 0.8 to 3.9 kpc. They are distributed non-uniformly along the galactic plane with longitude ranging from about 15 to 290 deg. The cluster size ranges from $\sim$ 1 to 10 pc with an average radius of $\sim$ 4.5 pc, while the galacto-centric distance ranges from $\sim$ 6 to 12 kpc. Eight clusters of the sample have UBVJHK data while remaining six have UBVRIJHK data. In those cases in which the RI photometry was reported in the Kron-Cousin system, the colours are converted to Johnson's VRI system using the relations given by Bessell (1979). We have used homogeneous and accurate data if they are available from more than one sources. The observations used in this present analysis are mostly based on modern optical and infrared imaging except for NGC 1502, Tr 15 and Tr 37. They have accuracies generally better than 0.02 mag in V, (B$-$V),
(V$-$R) and (V$-$I) and 0.03 mag in (U$-$B). The accuracy of JHK data are generally $\sim$ 0.07 mag.
 The spectroscopic data
wherever available are also taken. The accuracy of MK classification is generally better than two subclasses in spectral type and a class in luminosity.
\begin{figure}
\psfig{file=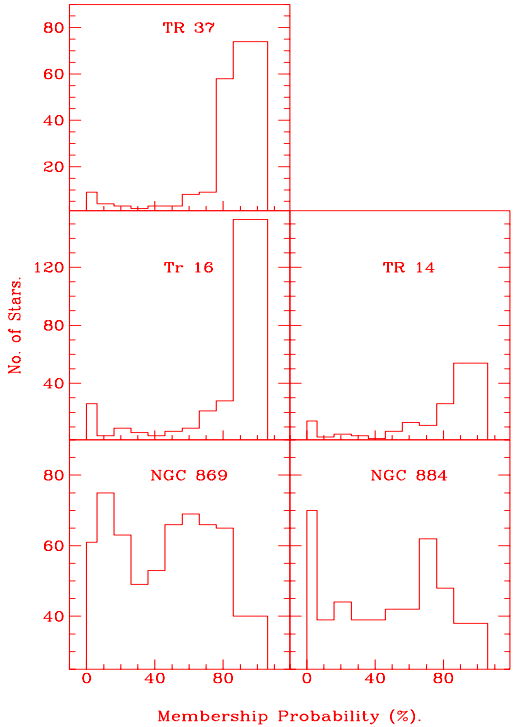,width=12cm,height=10cm}
\caption{Proper motion membership probability distribution of stars in the regions of NGC 869, NGC 884, Tr 14, Tr 16 and Tr 37. The source of data is indicated in Table 1.}
\end{figure}

\subsection{Cluster Membership}
Selection of cluster members is important for the study of interstellar extinction across the cluster region. Among the most commonly used methods, the one based on kinematical data (proper motion and radial velocity) is considered the most reliable. As reliable proper motion data are available for 8 of the 14 clusters under study, we use them to select cluster members.
The histograms of the membership probability (P) of all the investigated stars in clusters NGC 869, NGC 884, Tr 14, Tr 16 and Tr 37 are shown in Fig. 1. For NGC 2244, NGC 2264 and NGC 6611 such diagrams are available in the sources listed in Table 1. The histograms of P distribution indicate that in a cluster, stars are generally either in the low P group ( $<20\%$)
 or high P group ($>50\%$). It means that the proper motion studies of
these regions have been able to
segregate cluster members from field stars. The probability of inclusion of field stars in the sample is reduced if stars with P$>50\%$ are considered as cluster members and the same has been done here. For the five clusters NGC 1502, NGC 1893, Coll 228, Tr 15 and Be 86 the photometric criteria have been used to segregate cluster members from field stars. For NGC 663, photometric criteria has been applied to the sample of stars having P$\ge$1$\%$, as the proper motion data alone are unable to separate the cluster members from field stars.
\section{Determination of colour excesses}
To study the behaviour of the reddening and extinction law across the cluster region, we derive $E(U-V)$, $E(B-V)$, $E(V-R)$, $E(V-I)$,
$E(V-J)$, $E(V-H)$, $E(V-K)$ colour excesses of the cluster members. This was accomplished by comparing
the observed colours of the stars with their intrinsic colours derived from the
MKK spectral type-luminosity class colour relation given by FitzGerald (1970)
for (U$-$V) and (B$-$V); by Johnson (1966) for (V$-$R) and (V$-$I); and by Koornneef
(1983) for (V$-$J), (V$-$H) and (V$-$K). MK classification based on relatively higher
dispersion was preferred in cases where
it is available from more than one sources.
 In order to have a statistically significant number
of sample stars, photometric spectral types are determined using (U$-$B) and (B$-$V) data and the Johnson $\&$ Morgan (1953) Q - method (see Sagar $\&$ Joshi 1979 for details) for those stars whose positions in the (U$-$B), (B$-$V) and V, (B$-$V) diagrams of the clusters (given in the literature) indicate that they are early type MS stars. In the clusters, the number of stars used in the present analysis are given in Table 1. As the median P values of the cluster members range from  about 70 to 95$\%$, expected number of field stars in the sample range from 5 to 30$\%$ (see Table 1). The presence of some non-members in the sample may therefore not affect the conclusions of the present analysis. Wherever possible, homogeneous photometric data have been used. All these were done to increase the accuracy of colour excess estimation.
$E(U-V)$, $E(B-V)$, $E(V-R)$ and $E(V-I)$ are generally uncertain by $\sim$0.08 mag, while $E(V-J)$, $E(V-H)$ and $E(V-K)$ have typical errors of $\sim$0.15 mag. This is mainly because UBVRI magnitudes are more accurate than JHK magnitudes.

\section{Presence of non-uniform extinction}
To see the extent of non-uniform extinction in different clusters under study,  we plot the histograms of $E(B-V)$ in Fig. 2. This indicates that clusters have a different amount of non-uniform reddening as there is a wide range in the $E(B-V)$ values of cluster members.

        To see the presence of non-uniform extinction in a cluster, we calculate the value
of $\Delta$$E(B-V)$ = $E(B-V)$$_{max}$ $-$ $E(B-V)$$_{min}$, where $E(B-V)$$_{max}$ and $E(B-V)$$_{min}$ are determined on the basis of, respectively, the five highest and five lowest $E(B-V)$ values of the MS cluster members. The values of $\Delta$$E(B-V)$ obtained in this way along with other relevant informations are listed in Table 2.
 Apart from the non-uniform extinction, the other factors
 for the observed dispersion in $E(B-V)$ are stellar evolution, stellar duplicity, stellar
rotation, difference in chemical composition, dispersion in ages, dispersion in
distances, presence of non member stars and inaccuracies in the photometric data
(cf. Burki 1975, Sagar 1987). However, they (excluding the differential extinction)
 can produce a maximum dispersion in $\Delta$$E(B-V)$ of about 0.0 - 0.11 mag
for MS stars. So,
 a value $\Delta$$E(B-V)$$>$ 0.11 mag for a cluster is considered as an indication of the presence of non-uniform
 gas and dust in the direction of the clusters. This conclusion is also supported by near-infrared
 photometric studies of young open star clusters (cf. Tapia et al. 1988; Roth 1983; Sagar
 $\&$ Qian 1989). Since $\Delta$$E(B-V)$ values for all the clusters under study range from 0.22 to 1.03 mag, they indicate the presence of non-uniform extinction in the direction of the clusters. We thus confirm the results obtained earlier by other
 investigators, e.g. Tapia et al. (1991) in NGC 663;
Tapia et al. (1984) in NGC 869 and NGC 884; Vallenari et al. (1999) in NGC 1893 and Be 86;
 Per\'{e}z, Th\'{e} $\&$ Westerlund (1987) in NGC 2244 and NGC 2264; Sagar (1987) and Hillenbrand et al. (1994) in NGC 6611
and Tapia et al. (1988) in Tr 14, Tr 15, Tr 16 and Coll 228. The presence of non-uniform extinction has been indicated in NGC 1502 and Tr 37 for the first time. In order to understand the possible reasons for non-uniform extinction in the young star clusters, we carried out the following studies
\begin{figure}
\psfig{file=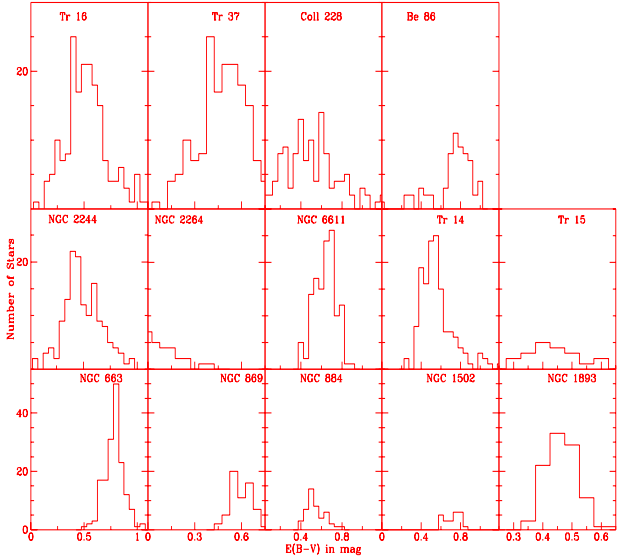,width=8.5cm,height=10cm}
\caption{The presence of non-uniform extinction in all clusters under study indicated as the dispersions in $E(B-V)$ are significantly larger than that expected from other sources (see text).}
\end{figure}
\begin{table*}
 \centering
  \begin{minipage}{140mm}
   \caption{The area investigated around central coordinates of the clusters under study are listed. The values of mean $E(B-V)$, $E(B-V)$$_{min}$, $E(B-V)$$_{max}$ and $\Delta$$E(B-V)$ are also given.}

 \begin{tabular}{lcccccccl}
\hline
&&&&&&\\
Cluster& Central Coordinates& Area& $\overline{E(B-V)}$& $E(B-V)$$_{min}$& $E(B-V)$$_{max}$& $\Delta$$E(B-V)$\\
&$\alpha_{2000}$~~~~~~~~~~~~$\delta_{2000}$&(arc min)$^{2}$&(mag)&(mag)&(mag)&(mag)\\
\hline
NGC 663&~~01$^{h}$46$^{m}$03$^{s}$~~~~~~$61^{\circ}15^{\prime}$.0&~~~~36&0.71&0.59&1.02&0.43\\
NGC 869&~~02~~19~~02~~~~~~~57~~08.8&~~~~24&0.61&0.51&0.73&0.22\\
NGC 884&~~02~~22~~27~~~~~~~57~~06.5&~~~~24&0.57&0.44&0.74&0.30\\
NGC 1502&~~04~~07~~43~~~~~~~62~~20.0&~~200&0.74&0.62&0.84&0.22\\
NGC 1893&~~05~~22~~41~~~~~~~33~~23.8&1300&0.49&0.39&0.63&0.24\\
NGC 2244&~~06~~32~~21~~~~~~~04~~51.7&2800&0.54&0.14&0.89&0.75\\
NGC 2264&~~06~~41~~03~~~~~~~09~~53.1&~~400&0.13&0.02&0.31&0.29\\
Tr 14&~~10~~43~~56~~~~~$-$59~~33.7&~~960&0.59&0.33&1.10&0.77\\
Tr 15&~~10~~44~~45~~~~~$-$59~~21.7&~~~~45&0.46&0.35&0.57&0.22\\
Tr 16&~~10~~45~~08~~~~~$-$59~~42.7&~~960&0.55&0.16&1.06&0.90\\
Coll 228&~~10~~43~~01~~~~~$-$60~~00.7&&0.54&0.12&1.15&1.03\\
NGC 6611&~~18~~18~~50~~~~~$-$13~~46.7&1600&0.66&0.43&0.87&0.44\\
Be 86&~~20~~20~~26~~~~~~~38~~41.5&~~~~36&0.74&0.24&1.01&0.77\\
Tr 37&~~21~~39~~03~~~~~~~57~~29.6&~~~~40&0.52&0.39&0.67&0.28\\
\hline
\end{tabular}
\end{minipage}
\end{table*}
\begin{figure}
\psfig{file=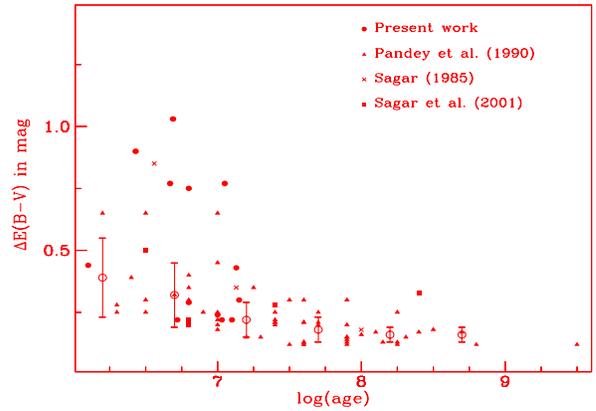,width=9cm,height=8cm}
\caption{Variation of $\Delta$$E(B-V)$ ($\equiv$ $E(B-V)$$_{max}$ - $E(B-V)$$_{min}$) with cluster age. The values of $\Delta$$E(B-V)$ are taken from different sources as indicated in the plot. The open circle and associated error bar denote the mean and standard deviation of the binned data (see text).}
\end{figure}
\subsection{Variation of $\Delta$$E(B-V)$ with age}
Young star clusters (age $\le10^{7}$ yr) are embedded in dust
and gas clouds and it is assumed that with time, gas and dust in these clouds
 will either be used up in star formation processes or will be dispersed
 by radiation pressure due to massive stars present in these systems or both. Therefore
in young open clusters, the presence of a variable amount of unused gas and dust
is expected to cause non-uniform extinction. In order to verify this, we plot in Fig. 3 cluster age against $\Delta$$E(B-V)$ which is considered as representative of non uniform extinction. The values of $\Delta$$E(B-V)$ for the clusters are taken from the present work (see Table 2); Sagar (1985); Sagar, Munari $\&$ de Boer (2001) and Pandey, Mahra $\&$ Sagar (1990) while the corresponding ages are taken from the respective sources if available. Otherwise, they are adopted from Mermilliod (1995). In order to see the general variation of mean $\Delta$$E(B-V)$ with age, we binned the data in various age groups and plot their mean $\Delta$$E(B-V)$ and standard deviation in Fig. 3. This shows that there is a variation of $\Delta$$E(B-V)$ with age of the cluster. In clusters older than $10^{8}$ years, there seems to be no signature of the presence of gas and dust inside the clusters, as the scatter is minimal. A large scatter in $\Delta$$E(B-V)$ is found in clusters younger than 10 Myr indicating the presence of varying amounts of gas and dust inside them. The observed relation indicates that with time, primordial gas and dust are either being used in star formation or being blown away by the hot stars present in the clusters, or both, as expected.

\subsection {\bf Spatial variation of colour excess $E(B-V)$}

    To study the spatial variation of colour excess across the cluster region, we
divide the cluster field into equal areas of small boxes of size $5^{\prime}$x5$^{\prime}$ for the clusters
NGC 663, NGC 869, NGC 884, NGC 1893, Tr 14, Tr 15 and Be 86 of the sample. For the clusters NGC 2244, NGC 2264,
and NGC 6611 such a study has already been carried out by Ogura $\&$ Ishida (1981), Sagar
 $\&$ Joshi (1983) and Sagar $\&$ Joshi (1979) respectively. The spatial variation of $E(B-V)$ could not be studied in Coll 228 as the cluster members occupy a small area on the sky while in NGC 1502, Tr 15 and Tr 37, statistically insignificant number (see Table 1) of cluster members denied such a study. Except for Tr 14, there is no systematic correlation of colour excess with position implying that gas and dust responsible for variable extinction  may be distributed randomly within the cluster. Similar behaviour has also been found by Sagar (1987) in young clusters NGC 654, NGC 2264, NGC 6823, NGC 6913, IC 1805, NGC 6530 and NGC 6611. In Tr 14, $E(B-V)$ appears to vary systematically with position (see Table 3). The reddening increases from east to west in the southern part of the cluster. The clusters NGC 663, NGC 869, NGC 884, NGC 1893, Tr 15 and Be 86 do not show any positional variation of E(B-V).
 
\begin{table*}
 \centering
 \begin{minipage}{140mm}
 \caption{Spatial variation of $E(B-V)$ across the cluster Tr 14. The mean values of $E(B-V)$ with their standard deviation in mag in 5$^{\prime}$ x 5$^{\prime}$ areas are indicated in the appropriate boxes, with the number of stars used for this purpose given in brackets. Coordinates are relative to the cluster centre given in Table 2.}

\footnotesize
\vspace{0.5cm}
\begin{tabular}{|ccccccc} \hline
&&&&&&           \\
$\Delta\alpha$ $\rightarrow$  &$-$35 to $-$30&$-$30 to $-$25&$-$10 to $-$5&$-$5 to 0&0 to 5&5 to 10\\
$\Delta\delta$~~ &&&&&&        \\
$\downarrow$ &&&&&&  \\
\hline
&&&&&&           \\
$-$5~ to 0&-&-&0.73&0.72$\pm$0.3&0.54$\pm$0.7&0.47$\pm$0.2\\
&&&(1)&(10)&(12)&(4)\\
&&&&&& \\
0 to 5&-&-&0.82$\pm$0.22&0.56$\pm$0.9&0.58$\pm$0.2&0.44$\pm$0.05\\
&&&(3)&(15)&(14)&(4)\\
&&&&&& \\
10 to 15&0.71$\pm$0.10&0.59&-&-&-&-\\
&(4)&(1)&&&&\\
&&&&&& \\
\hline
\end{tabular}
\end{minipage}
\end{table*}

\subsection{Variation of colour excess $E(B-V)$ with luminosity}
In order to study the variation of colour excess with luminosity, we have
 plotted colour excess $E(B-V)$ against absolute magnitude $M_{V}$ for
the clusters under study in Fig. 4, except for NGC 2244, NGC 2264 and NGC 6611 since for them such plots are provided by Sagar (1987). In order to convert apparent V values into $M_{V}$ we use the relation $A_{v}$ = 3.25$E(B-V)$ and distances given in Table 1. We also plot in the figure the variation of mean $E(B-V)$ and its standard deviation with $M_{V}$. For estimating these values we grouped cluster members in such a way that five or more stars are present in a group. An inspection of Fig. 4 indicates that:

\begin{figure}
\psfig{file=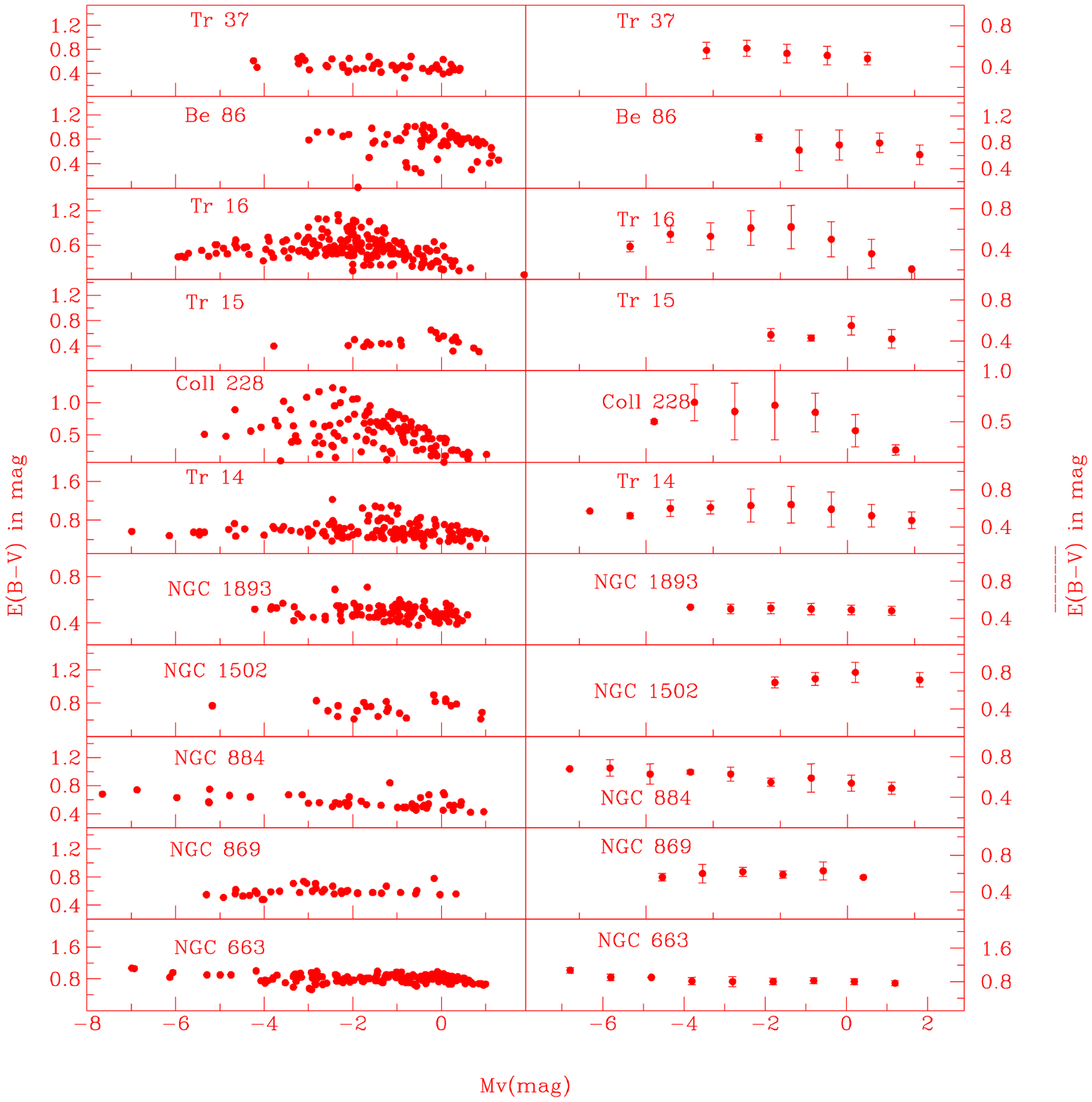,width=10cm,height=11cm}
\caption{$E(B-V)$ versus Mv diagrams for NGC 663, NGC 869, NGC 884, NGC 1502, NGC 1893, Tr 14, Coll 228, Tr 15, Tr 16, Be 86 and Tr 37. The data points are shown in the left panel while the right panel shows the plot of mean $E(B-V)$ and its standard deviation.}
\end{figure}
\begin{enumerate}
\item Clusters NGC 663, NGC 869, NGC 884, NGC 1502, NGC  1893, Tr 14, Tr 15 and Tr 37 show no correlation of colour excess with luminosity. The same behaviour was also found by Sagar (1987) for the clusters namely NGC 1976, NGC 2244, NGC 2264, NGC 4755 and NGC 6530.
\item In clusters Coll 228 and Tr 16, both mean values and scatter in $E(B-V)$ first increase up to $M_{V}$ $\sim$ -2.0 mag and then they decrease with the $M_{V}$ of clusters members. The rate of increase in $E(B-V)$ with $M_{V}$ is considerably slower than the rate of decrease. The correlation between $E(B-V)$ and $M_{V}$ for these clusters is thus complex. 
\item In cluster Be 86, there is no variation of $E(B-V)$ with $M_{V}$ for $M_{V}$ $\le$ 0.0 mag, while for fainter members the values of $E(B-V)$ decrease with $M_{V}$.
\end{enumerate}

\subsection{Variation of colour excess $E(B-V)$ with spectral type}
Fig. 5 shows the variation of $E(B-V)$ against spectral class of the cluster members. In all the clusters we have only early (O, B and A ) type stars. Clusters NGC 663, NGC 869, NGC 884, NGC 1502, NGC 1893, Tr 14, Tr 15 and Tr 37 show no correlation of $E(B-V)$ with spectral type. However, for Coll 228, Tr 16 and Be 86, $E(B-V)$ decreases as one moves from B to A spectral type, while it shows no such variation in O type stars. Sagar (1987) also found no corelation of $E(B-V)$ with spectral class in the clusters NGC 1976, NGC 2244, NGC 2264, NGC 4755, NGC 6530 and NGC 6611 but in the case of extremely young clusters NGC 6823, NGC 6913 and IC 1805, a decreasing trend of $E(B-V)$ with spectral class was observed.

 The mean value of $E(B-V)$ and its standard deviation with spectral type are also plotted in Fig. 5 for studying the variation of scatter in $E(B-V)$ with spectral type. For estimating these values we grouped cluster members of one subspectral type.  No significant trend is seen between the scatter in $E(B-V)$ and spectral class.

\subsection{Discussion on non-uniform extinction}

On the basis of Figs. 4 and 5 we can conclude that NGC 663, NGC 869, NGC 884, NGC 1502, NGC 1893, Tr 14, Tr 15 and Tr 37  do not show any variation of $E(B-V)$ with luminosity as well as spectral class. The clusters Coll 228, Tr 16 and Be 86 show a complex correlation of the $E(B-V)$ with luminosity as well as spectral class. In these clusters, $E(B-V)$ decreses with spectral type and $M_{V}$ for the B stars, while O stars have higher reddening than the B stars. The correlation of $E(B-V)$ against $M_{V}$ for Tr 16 and Coll 228 reaches a maximum near $M_{V}$ = -2.0 mag, then decreases for brighter stars.  

     The dependency of colour excess $E(B-V)$ with luminosity and spectral class for the clusters Coll 228, Tr 16 and Be 86 can be explained with the help of theories of star formation. According to Krelowski $\&$ Strobel (1981), highly reddened luminous stars  in association and young clusters may have relic circumstellar matter around them. On the other hand, theoretical  models for star formation processes given by Yorke $\&$ Krugel (1976) and Bhattacharjee $\&$ Williams (1980) suggest that the relative mass of the relic envelope present around  newly formed massive stars is positively correlated with the stellar mass. MS stars emit strong stellar winds and UV radiation. The time required for them to blow off their relic material will depend upon factors like the mass of their relic envelope, resistance offered by the immediate surroundings, etc. (see Sagar 1987 for details).  Hence one may say that during star formation processes in molecular clouds, conditions around MS stars of the clusters Coll 228, Tr 16 and Be 86 are such, that relic circumstellar material is still present around fainter ($M_{v}$ $>$ -2.0 mag) stars, while it has been blown away from the brighter stars of these clusters. In addition to this all members in clusters where the $E(B-V)$ values are not dependent either on luminosity or spectral class, are either younger or have ages comparable to Coll 228, Tr 16 and Be 86.
\begin{figure}
\psfig{file=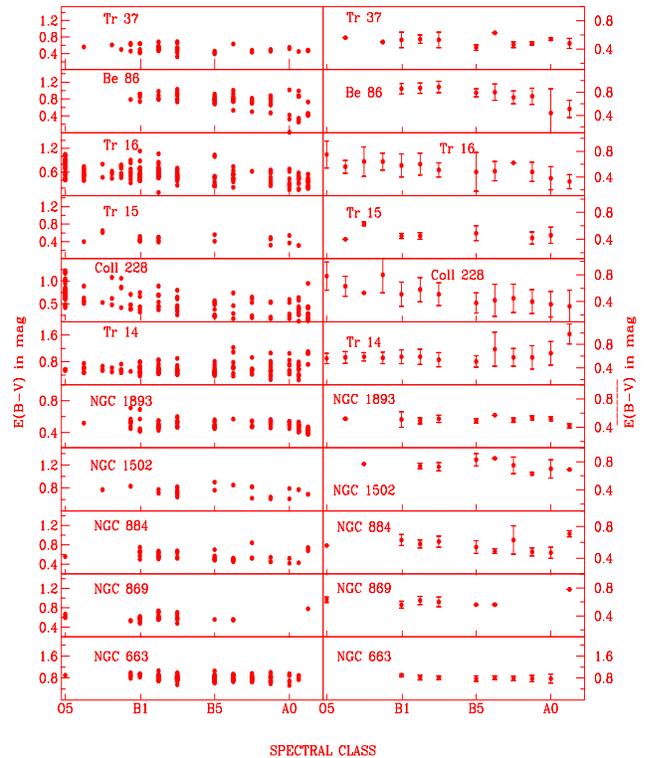,width=9.5cm,height=11cm}
\caption{$E(B-V)$ versus spectral class plots for the clusters NGC 663, NGC 869, NGC 884, NGC 1502, NGC 1893, Tr 14, Coll 228, Tr 15, Tr 16, Be 86 and Tr 37. The data points are shown in the left panel while the right panel shows the plot of mean $E(B-V)$ and its standard deviation.}
\end{figure}
\begin{figure*}
\psfig{file=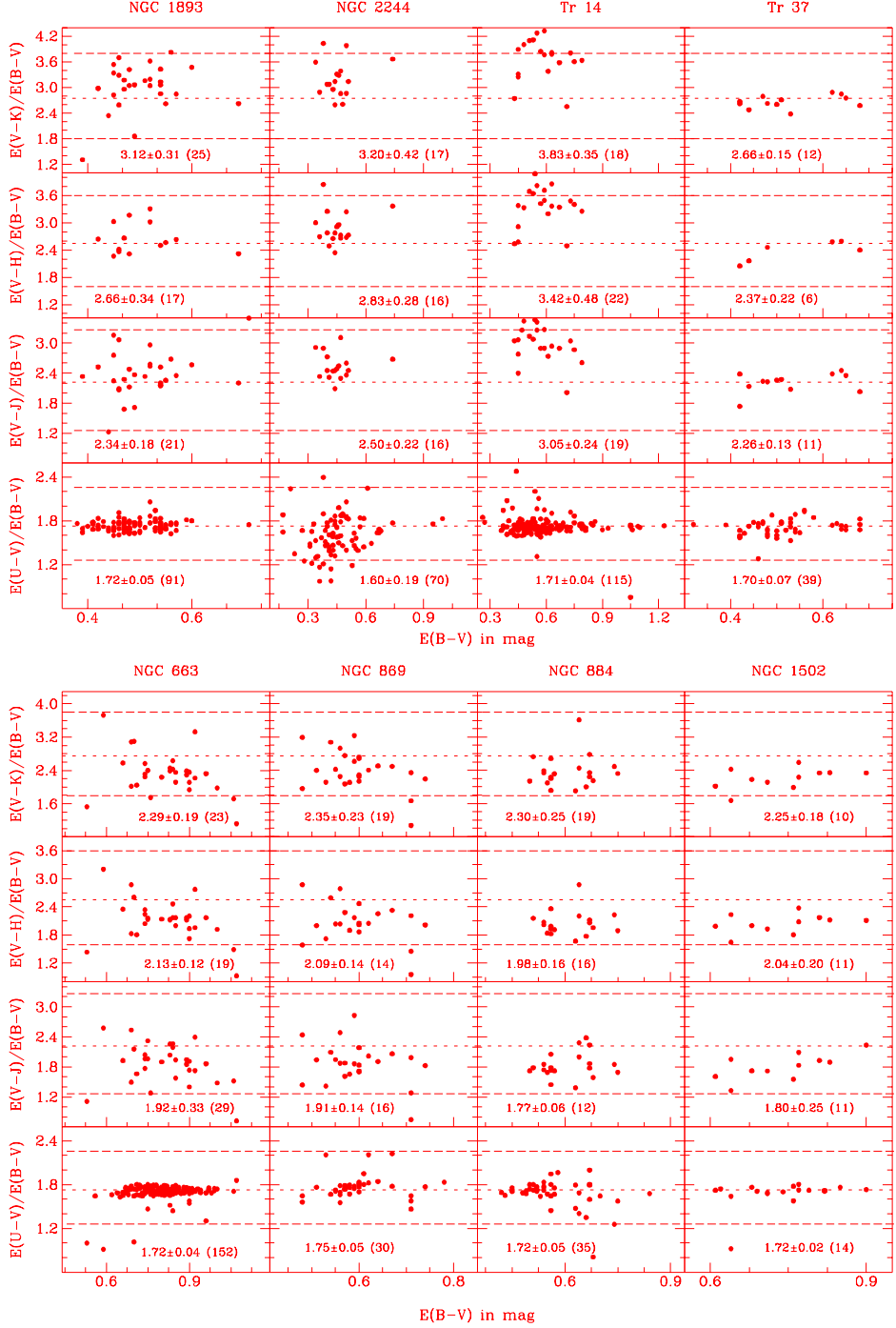,width=19cm,height=19cm}
\caption{The colour excess ratios $E(U-V)$/$E(B-V)$, $E(V-J)$/$E(B-V)$, $E(V-H)$/$E(B-V)$ and $E(V-K)$/$E(B-V)$ are plotted against $E(B-V)$ for NGC 663, NGC 869, NGC 884, NGC 1502, NGC 1893, NGC 2244, Tr 14 and Tr 37. The dotted lines show the ratio for a normal interstellar extinction law along with the boundaries expected from observational uncertainties of 30$\%$ in the optical and 40$\%$ in the near-IR (dashes lines). The mean values of the observed colour excess ratios along with their standard deviation and number of stars used in deriving them (inside the bracket) are shown on the plot.}
\end{figure*}
\begin{figure*}
\psfig{file=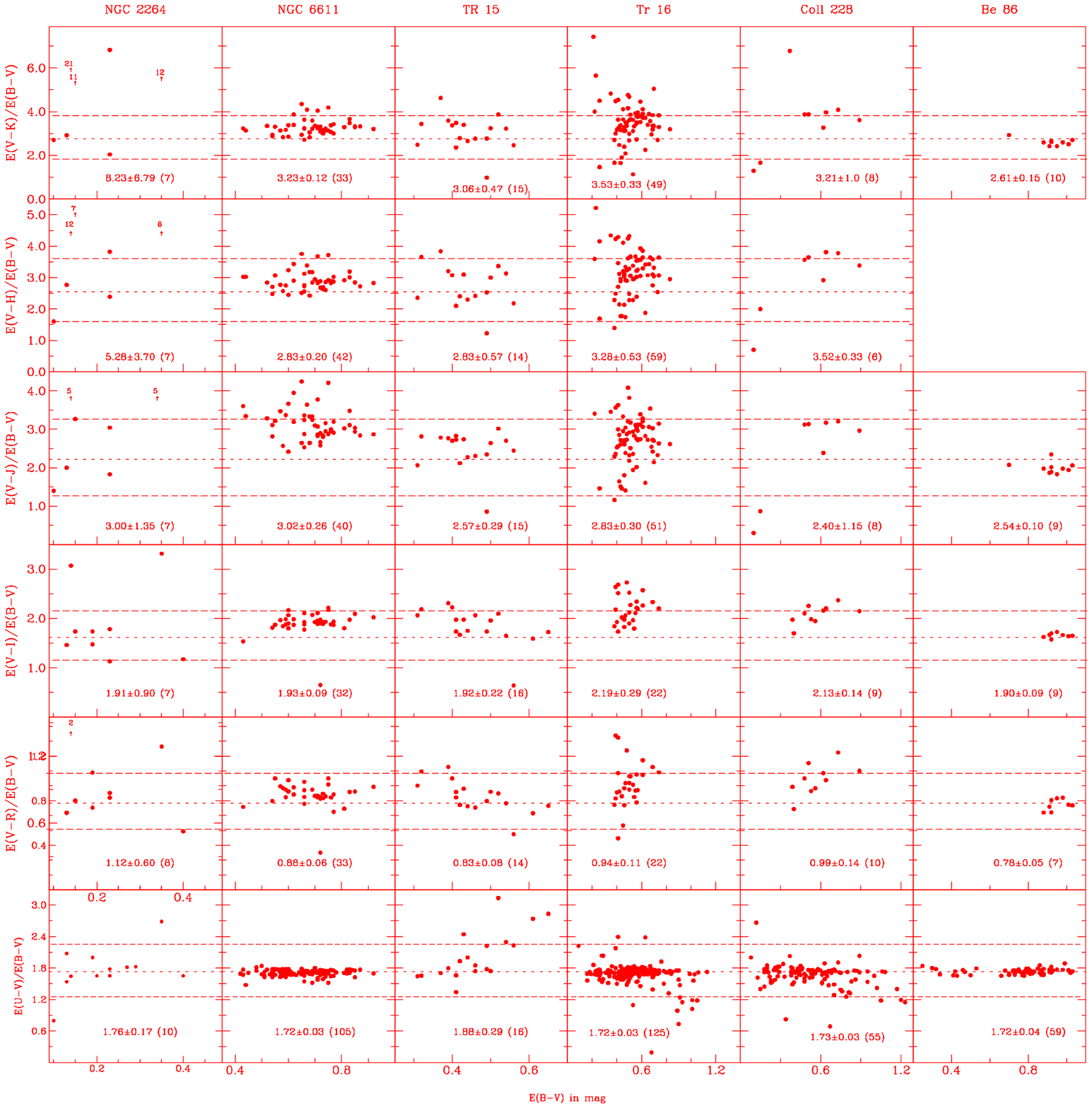,width=19cm,height=19cm}
\caption{The colour excess ratios $E(U-V)$/$E(B-V)$, $E(V-R)$/$E(B-V)$, $E(V-I)$/$E(B-V)$, $E(V-J)$/$E(B-V)$, $E(V-H)$/$E(B-V)$ and $E(V-K)$/$E(B-V)$ are plotted against $E(B-V)$ for NGC 2264, NGC 6611, Tr 15, Tr 16, Coll 228 and Be 86. The dotted and short dashed lines and symbols are the same as in Fig. 6.}
\end{figure*}
\section{Extinction Law}
\noindent In order to understand the properties of interstellar matter causing non-uniform extinction in the direction of clusters, we plot the ratios of colour excesses $E(U-V)$, $E(V-J)$, $E(V-H)$ and $E(V-K)$ with $E(B-V)$ against it in Fig. 6 for NGC 663, NGC 869, NGC 884, NGC 1502, NGC 1893, NGC 2244, Tr 14 and Tr 37. The ratio of colour excesses $E(U-V)$, $E(V-R)$, $E(V-I)$, $E(V-J)$, $E(V-H)$ and $E(V-K)$ with $E(B-V)$ for the members of NGC 2264, NGC 6611, Coll 228, Be 86, Tr 15 and Tr 16 are plotted against $E(B-V)$ in Fig. 7. We have considered only those cluster members which have $E(B-V)$ $>$ 0.1 mag, as colour excess ratios for stars with smaller $E(B-V)$ values have large uncertainties. This criterion has excluded only a few stars in NGC 2264 since for all other clusters under study $E(B-V)$$_{min}$ is $>$ 0.1 mag (see Table 2). In Fig. 6 and 7, horizontal dotted lines represent the colour excess ratios corresponding to the normal interstellar extinction law described by Mathis (1990). Errors expected from the observational uncertainties in the colour excess ratios are also shown as short dashed horizontal line in the figures. They are $\sim$ 30$\%$ in the optical but become larger, $\sim$ 40$\%$, in the near-IR. The average values of the observed colour excess ratios along with its $\sigma$ and number of stars used in determining them are also given in the appropriate boxes. An inspection of Figs. 6 and 7 clearly indicates that:- 

\begin{enumerate}

\item The average value of $E(U-V)$/$E(B-V)$ for all the clusters agree within 1$\sigma$, where $\sigma$ is the standard deviation of data points, with the normal interstellar value of 1.73. However, there are a few stars in NGC 663, NGC 884, NGC 1502, NGC 2244, NGC 2264, Tr 14, Tr 15, Tr 16 and Coll 228 showing ratios that deviate more than their observational error, but most are within twice their errors. In Tr 15, all colour excess ratios are not too different from the normal one except for $E(U-V)$/$E(B-V)$ where values between 2.22 and 3.13, compared to the normal value of 1.73, are obtained for stars with $E(B-V)$ $>$ 0.45. Tapia et al. (1988) also noticed this anomaly. The least square linear regression to the data points yields $\frac{E(U-V)}{E(B-V)}$ = (3.82$\pm$0.88)$E(B-V)$ + (0.27$\pm$0.33). Tapia et al. (1988) conclude that Tr 15 is located in a denser intracluster dust cloud whose grains have been 'processed' in such a way as to produce disproportionally larger $E(U-V)$ excess per unit column density as compared to the average interstellar medium. In all the clusters, attenuating material seems to have the properties of normal interstellar matter except for a few stars whose anomalous extinction arises from their being a field star or a cluster star with peculiar intrinsic colours.

\item In the clusters NGC 663, NGC 869, NGC 884 and NGC 1502 the colour excess ratios for members at $\lambda$ $\ge$ $\lambda_{J}$ are generally smaller than the normal values,
when compared with $E(B-V)$ (see Fig. 6). This and the values of R (=1.1$E(V-K)$/$E(B-V)$ following Whittet $\&$ van Breda (1980)) of these clusters indicate that the size distribution of the dust particles in the direction of these clusters is biased towards smaller values than the size of normal interstellar matter such that they produce smaller values of colour excess ratios when compared with $E(B-V)$, but normal values while compared with $E(V-J)$. Cr\'{e}z\'{e} (1972) also indicates
that in the direction of all these clusters, the value of R is relatively
lower than other sight lines through the Galaxy. Tapia et al. (1991) have also shown that the size distribution of dust particles in the direction of the cluster NGC 1502 is biased towards smaller values. All these clusters have a similar age ($\sim$ 12 Myr) and are located in a small sector of the Galactic plane with $130^{o}$ $<$ l $<$ $145^{o}$. We therefore conclude that the interstellar matter in this part of the Galaxy is somewhat different from the normal one.

\item The colour excess ratios of the members in NGC 1893, NGC 2244, Tr 37 and Be 86 seems to be generally normal at all wavelengths from U to K passbands (see Fig. 6 and 7).
  This indicates that the extinction law is close to normal galactic extinction in the direction of these  clusters. Per\'{e}z et al. (1987) and Morbidelli et al. (1997
) have also shown that the value of R is normal in the direction of NGC 2244
and Tr 37 respectively. In the direction of NGC 1893, Tapia et al. (1991)
have also indicated an average galactic extinction law.

\item The colour excess ratios to $E(B-V)$ of the cluster members are larger than the normal values at $\lambda$ $\ge$ $\lambda_{R}$ in NGC 6611, Coll 228 and Tr 16; and at $\lambda$ $\ge$ $\lambda_{J}$ in Tr 14, but the values become normal when compared to $E(V-J)$ except in NGC 6611 where it is slightly lower than the corresponding normal values. This implies that the extinction law is anomalous in the sense that, in these regions, grain size distribution seems to be biased towards larger than normal size particles. Hillenbrand et al. (1993) have also reported a higher value of R for NGC 6611.

\item The cluster NGC 2264 exhibits a large scatter from the normal values of the colour excess ratios. Per\'{e}z et al. (1987) also found a large deviation in the value of R most probably due to the presence of circumstellar material around some cluster members (see next section).
\end{enumerate}

\begin{figure}
\psfig{file=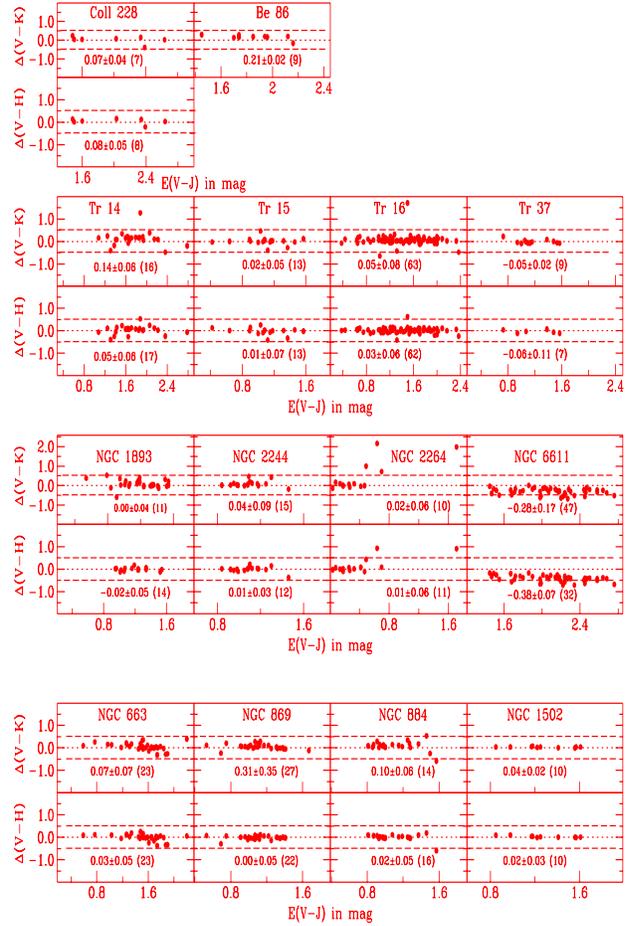,width=8.5cm,height=13cm}
\caption{Plots of near-IR flux excess/deficiency in terms of $\Delta$(V-H) and $\Delta$(V-K) against the colour excess $E(V-J)$ (see text). The horizontal dotted lines denote zero excess.The short dashed lines denote the extent of the expected errors.}
\end{figure}
\section{Near-IR Fluxes}
 \begin{table*}
 \centering
 \begin{minipage}{140mm}
 \caption{Near-IR flux excess/deficiency in members of clusters under study.}
\begin{tabular}{lcccccccl}
\hline
cluster&Webda Star No.&V&$E(V-J)$ &$\Delta$(V$-$H)&$\Delta$(V$-$K)&Sp Type \\
&&(mag)&(mag)&(mag)&(mag)&\\
\hline
NGC 884&1781&9.26&1.57&$-$0.60~~~&$-$0.58~~~&B1IV\\
NGC 2264&90&12.71&1.17&0.91&1.98&B4V\\
&100&10.04&0.64&0.93&2.16&A2IV\\
&165&10.99&0.49&0.43&1.00&A2V\\
&46&9.18&0.70&0.09&0.72&A3V\\
Tr 14&15&11.85&1.88&0.53&1.27&B7V\\
Tr 16&68&12.14&1.50&0.63&1.71&O5V\\
\hline
\end{tabular}
\end{minipage}
\end{table*}

Contributions towards the reddening measured in terms of colour excesses of the cluster members include (i) interstellar extinction between the observer and the cluster, (ii) intra-cluster extinction due to the gas and dust present inside the cluster; and (iii) circumstellar extinction due to remnant disks or envelopes, which have not got fully dissipated from the cluster members because of their youth. If one assumes that the interstellar material in the foreground of the clusters follows the normal interstellar extinction law, then whatever deviations are seen from the normal values in the colour excess ratios of stars is only due to (ii) or (iii) or both. The colour excess used to measure interstellar matter should, however be selected carefully as the clusters under study contain young stellar objects. In such circumstances one uses $E(V-J)$ instead of $E(B-V)$ (Smith 1987; Tapia et al. 1988,1991; Sagar and Qian 1989, 1990), because $E(V-J)$ does not depend on environmental properties such as chemical composition, shape, structure and the degree of alignment of the interstellar matter (Vashchinnikov $\&$  I1'in 1987; Cardelli, Clayton $\&$ Mathis 1989). The differences between the observed colour excess in (V$-$H) and (V$-$K) based on spectral classification and the derived colour excess from $E(V-J)$ assuming normal extinction law are calculated. These differences are plotted with $E(V-J)$ in Fig. 8.  The major sources of error in the differences are the observational uncertainties in JHK magnitudes, inaccuracies in estimation of $E(V-J)$ and errors in the spectral classification. The differences can therefore be considered statistically significant only if their absolute values are larger than $\sim$0.5 mag.

An inspection of Fig. 8 indicates that the values of $\Delta$(V$-$H) and $\Delta$(V$-$K) are distributed randomly around the horizontal line, except in the case of NGC 6611, where they are systematically lower. Most of the cluster members have absolute values of $\Delta$(V$-$H) and $\Delta$(V$-$K) close to zero indicating that there are no near-IR excess fluxes. However, a few stars show an excess and deficiency in radiation at H and K wavelengths. They are listed in Table 4. We can clearly see from this table that all the stars are producing excess flux more in K compared to H, except one star NGC 884-1781 which has approximately equal flux deficiency in both bands. Strom et al. (1971, 1972) noticed circumstellar envelope around star NGC 2264-90. Sung, Bessell $\&$ Lee (1997) identified NGC 2264-100 as a pre-main sequence star, while Young (1978) found $H_{\alpha}$ emission in star NGC 2264-64. Smith (1987) found an infrared excess in star Tr 14-15. These observations, thus support our results listed in Table 4. We have identified for the first time the presence of a near-infrared excesses in NGC 2264-165 and Tr 16-68 and a deficiency in NGC 884-1781. The stars are of O, B and A type main sequence, except NGC 884-1781 and NGC 2264-100.

\section{Conclusions}

\noindent From the present analysis of reddening and ratios of colour excesses of members in 14 young (age $\le$ 20 Myr) open clusters, the
following main conclusions can be drawn:                                       
 
\begin{enumerate}
 \item  Non-uniform extinction decreases systematically with the age of the
clusters. We therefore conclude that young open star clusters
(age $\le10^{7}$ yr) still have unused gas and dust inside them, while in older
 clusters (age $\ge10^{8}$ yr), they are either used in star formation or blown away by the UV radiation and strong winds of hot stars or both.
\item  The cluster members of Coll 228, Tr 16 and Be 86 show significant variation of $E(B-V)$ with luminosity and spectral class. This suggests that some stars have retained their circumstellar material whilst it has been stripped away from most others. In these clusters, the colour excess ratios at $\lambda$ $\ge$ $\lambda_{R}$ have larger values than normal. However, the presence of near-IR excess is not observed at $\lambda$ $\le$ 2.2 $\mu$m in most of the members of Coll 228, Tr 16 and Be 86 except for one star (see Table 4). Observations at longer wavelengths are required to confirm the presence or absence of circumstellar dust.
\item There is no significant variation of colour excess spatially except in Tr 14. In the clusters NGC 663, 869, 884, 1893, Tr 16 and Be 86, random variation of $E(B-V)$ is present. In the southern part of the cluster Tr 14, the value of $E(B-V)$ increases systematically as one moves from east to west (see Table 3).

\item Colour excess diagrams show that cluster members generally follow the normal interstellar extinction law at optical wavelengths, while anomalous behaviour has been noticed at $\lambda$ $\ge$ $\lambda_{J}$ for stars in some clusters (see Figs 6 and 7). In the direction of a small part of the Galaxy ( longitude ranging from $\sim$ 130 to 145 deg along the Galactic plane) the size distribution
of the dust particles is biased towards smaller values than normal.
\item Some stars show anomalous behaviour at H and K. The near-IR excess radiation, which generally increases with wavelengths from H to K could be due to the presence of hot dust close to the stars or perhaps a companion? In contrast, star NGC 884-1781 show the same near-IR flux deficiency in both H and K passband.
\end{enumerate}
On the basis of the above, one may conclude that there is no uniformity in extinction properties amongst these clusters. In fact, they differ from cluster to cluster. All this indicates that non-uniform extinction observed in young clusters cannot be understood in terms of a simple physical scenario. Actually, it depends upon factors such as the age of the cluster members, initial spatial distribution of matter in the molecular clouds, sequential star formation processes and the distribution of hot O and B stars in the cluster etc. A complicated physical scenario is therefore required to explain the non-uniform extinction in young star clusters.

\section*{Acknowledgments}

We are grateful to the referee, Dr. Mike Bessell, for valuable comments which improved the scientific content of the paper. Useful discussions with Dr. A. K. Pandey is thankfully acknowledged.

\bsp
\label{lastpage}


\begin{thebibliography}{99}
\bibitem{b1}Bessell M. S., 1979, PASP, 91, 589 
\bibitem{b2}Bhattacharjee S. K., Williams I. P., 1980, MNRAS, 192, 841 
\bibitem{b3}Bohannan B., 1975, AJ, 80, 625
\bibitem{b4} Burki G., 1975, A$\&$A, 43, 37
\bibitem{b5} Cardelli J. A., Clayton G. C., Mathis J. S., 1989, ApJ, 345,245
\bibitem{b6} Cudworth K. M., Martin S. C., Degioia-Eastwood K., 1993, AJ, 105, 1822
\bibitem{b7} Cr\'{e}z\'{e} M., 1972, A$\&$A, 21, 85
\bibitem{b8} FitzGerald M. P., 1970, A$\&$ A, 4, 234
\bibitem{b9} Hayashi C., 1970, Mem. Soc. R. Sci. Liege, 19, 127
\bibitem{b10} Johnson H. L., Moggan W. W., 1953, ApJ, 117, 313
\bibitem{b11} Johnson H. L., 1966, ARA$\&$A, 4, 193
\bibitem{b12} Hillenbrand A. L., Massey P., Strom S. E., Merrill K. M., 1993, AJ, 106, 1906
\bibitem{b13} Krelowski J., Strobel A., 1981, Acta Astr, 31, 313
\bibitem{b14} Koornneef J., 1983, A$\&$A, 128, 84
\bibitem{b15} Larson B., 1973 ARA$\&$A, 11, 219
\bibitem{b16} McNamara B. J., 1976, AJ, 81, 845
\bibitem{b17} Mermilliod J. \_C., 1995, in Egret, Abrecht M. A., eds, Information and on line data in Astronomy, Kulwar Academic Press, p. 227
\bibitem{b18} Marshall J., van Altena W., 1987, AJ, 94, 71
\bibitem{b19} Marshall J., van Altena W.F., Chiu L. T. G., 1982, AJ, 87, 1497
\bibitem{b20} Morbidelli L., Patriarchi P., Perinotto M., Barbaro G., Bartolomeo A.Di., A$\&$A, 327, 125
\bibitem{b21} Muminov M., 1983, BICDS, 24, 95
\bibitem{b22} Ogura K., Ishida K., 1981, PASJ, 33, 149
\bibitem{b23} Pandey A. K., Mahra H. S., Sagar R., 1990, AJ, 99, 617
\bibitem{b24} Per\'{e}z M. R., Th\'{e} P. S., Westerlund B. E., 1987, PASP, 99, 1050
\bibitem{b25} Mathis J. S., 1990, ARA$\&$A, 28, 37
\bibitem{b26} Reddish V. C., 1967, MNRAS, 135, 251
\bibitem{b27} Roth M., 1988, MNRAS, 233, 773
\bibitem{b28} Sagar R, 1985, Abastumani astrophys. Obser. Mt. Kanobili Bull., 59, 191
\bibitem{b29} Sagar R., 1987 MNRAS, 228, 483
\bibitem{b30} Sagar R, Qian Z. Y., 1989, MNRAS, 240, 551
\bibitem{b31} Sagar R, Qian Z. Y., 1990, ApJ, 353, 174
\bibitem{b32} Sagar R., Joshi U. C., 1979, Astrophys. Space Sci., 66, 3
\bibitem{b33} Sagar R., Joshi U. C., 1983, MNRAS, 205, 747
\bibitem{b34} Sagar R., Munari U., de Boer K. S., 2001, MNRAS (accepted)
\bibitem{b35} Smith R. G., 1987, MNRAS, 227, 943
\bibitem{b36} Strom K. M., Strom S. E., Yost J., Carrasco L., Grasdalen G., 1972, ApJ, 173, 353
\bibitem{b37} Strom K. M., Strom S. E., Yost J., 1971, ApJ, 165, 479
\bibitem{b38} Sung H., Bessell M. S., Lee S., 1997, AJ, 114, 2644
\bibitem{b39} Tapia M., Costero R. Echevarria, Roth M., 1991, MNRAS, 253, 649
\bibitem{b40} Tapia M., Roth M., Costero R., Navarro S., 1984, Rev. Mex. Astron. Astrofis., 9, 65
\bibitem{b41} Tapia M., Roth M., Marraco H., Ruiz M. T., 1988, MNRAS, 232, 661
\bibitem{b42} Vallenari A., Richichi A., Carraro G., Girardi L., 1999, A$\&$A, 349, 825
\bibitem{b43} Vashchinnikov N. V., I1'in V. B., 1987, SvA Lett., 13, 157
\bibitem{b44} Wallenquist A., 1975, Uppsala Astr. Obs. Ann. band, 5 no. 8
\bibitem{b45} Warner J. W., Strom S. E., Strom K. M., 1979, ApJ, 213, 427
\bibitem{b46} Whittet D. C. B., van Breda I. G., 1980, MNRAS, 192, 467
\bibitem{b47} Yorke H. W., Krugel E., 1976, A$\&$A, 54, 183
\bibitem{b48} Young A., 1978, PASP, 90, 144
\bibitem{b49} Zhao J., Tian K., Jing J., $\&$ Yin M., 1985, Special Issue for Tables of Membership for 42 open clusters. Shanghai Observatory, Akademia Sinica, Shanghai.
\end{thebibliography}
\end{document}